\documentclass[aps,twocolumn,showpacs,amssymb,%
floatfix,superscriptaddress]{revtex4}
\usepackage{graphicx}% Include figure files
\usepackage{epsf}
\usepackage{dcolumn}% Align table columns on decimal point
\usepackage{bm}% bold math
\usepackage{hyperref}
\usepackage{latexsym}

\begin{document}
\def\av#1{\langle#1\rangle}
\def\etal{{\it et al\/.}}
\def\pc{p_{\rm c}}
\def\l{{\lambda}}
\def\hm{h_*}
\def\xm{x_*}
\def\remark#1{{\bf *** #1 ***}}
\def\beq{\begin{equation}}
\def\eeq{\end{equation}}
% Include your paper's title here

\title{``Winner takes it all": strongest node rule for evolution of scale free
networks}
\author{H. \v Stefan\v ci\'c\thanks{shrvoje@thphys.irb.hr}}
\affiliation{Theoretical Physics Division, Rudjer Bo\v{s}kovi\'{c} Institute, \\
   P.O.Box 180, HR-10002 Zagreb, Croatia}
\author{V. Zlati\'c\thanks{vzlatic@irb.hr}}
\affiliation{Theoretical Physics Division, Rudjer Bo\v{s}kovi\'{c} Institute, \\
   P.O.Box 180, HR-10002 Zagreb, Croatia}

\begin{abstract}
We study a novel model for evolution of complex networks. We
introduce information filtering for reduction of the number of
available nodes to a randomly chosen sample, as stochastic component of evolution. 
New nodes are attached to the nodes that have maximal degree in the sample,
which is a deterministic component of network evolution
process. This fact is a novel for evolution of scale free networks
and depicts a possible new route for modeling network growth. We
present both simulational and theoretical results for network
evolution. The obtained degree distributions exhibit an obvious power-law
behavior in the middle with the exponential cut off in the end. This
highlights the essential characteristics of information filtering in the
network growth mechanisms.
\end{abstract}
\maketitle

\section{Introduction}

Recently, there have been a number of extensive investigations in the
field of complex networks. With such an extensive effort a number
of important theoretical and practical results have been
reported\cite{AB02,DM02,N03}. Many real world
systems can be described as complex networks: www\cite{www}, internet routers
\cite{internet1,internet2,internet3}, proteins \cite{Jeong01},
scientific collaborations \cite{GI95}, among others. The main features
that separate complex networks from ``ordinary" networks are the
famous small world effect \cite{WS98} and the scale free degree
distribution \cite{BA99b}. 

The first and simplest model for the scale free
distribution of degrees in complex network was proposed by Albert
and Barabasi \cite{BAJ99} (thereafter referred to as AB model). This
model is based on a simple principle of preferential attachment. The
network grows in such a way that at each time step $t$ a new
node is introduced into the network and attaches itself to some of
older nodes designated by the moment $s$ when they entered the network. The
probability that the node $t$ will attach 
itself to a node $s$ is linearly proportional to the degree $k_{s}$ of
the older node $P_{t\rightarrow s}\sim k_{s}$. Using this simple
principle a scale free network of exponent 3 is easily
reconstructed. Although very appealing because of its simplicity
the AB model cannot correctly reproduce all characteristics of real world networks.
First, it produces a temporally correlated network in the sense that older nodes
tend to have more edges than younger ones, which was not observed in real data
\cite{Lada00}. Second, it assumes that every new node has the
complete information about the whole network, which is unrealistic
for real network formations \cite{mossa,MI04a}. Third,
in its original form, it reproduces only networks with degree
distribution characterised by exponent $\gamma=3$. Nevertheless the AB model
has triggered a huge number of models that try to avoid
these shortcomings, but are also a natural extension of the original. Among
others there are models with nonlinear preferentiality
\cite{KRL00}, with rewiring of edges at later times \cite{AB00},
with a fitness parameter as an intrinsic value of a node
\cite{BB01a, BB01b}, etc. Although novel and more complex
approaches, that describe a variety of degree distributions and
have more support in the real data, have been studied recently
\cite{SBD04, CBM04}, we believe that it is also of fundamental
importance to examine ``as simple as possible" processes that
capture essential behavior of real world networks.

In this paper we present a novel model which exhibits power-law-like degree
distribution of an undirected network or the in-degree 
power-law-like distribution of a directed network. The purpose of
the model is to test information filtering as a stochastic
component of the network evolution process, while using a simple
deterministic rule for attachment of new nodes. The results we
 report in this article clearly show that our model can
reproduce power-law distributions but also power-laws with a
cut off, similarly to some real data reported recently
\cite{Batorski04}.

\section{Model}

Our model introduces two crucial features that make it different from the
Albert-Barabasi model. 
% The network growth is a standard one.
A new node is introduced into the network at each time step. For simulation
purposes, we first generate a network of 1100 nodes which are
completely randomly connected to each other. Each new node in this
core is connected to one of the older ones with uniform probability, until a
core is formed. The size of the core is taken to be 1100 because
we chose to monitor filtration subsets up to 1000 nodes. After the
core is formed, the following procedure takes place. Each new node
attaches itself to the network with $\omega$ links. To choose to
which of the already present nodes in the network it will attach itself, the
following rule is applied. i) A sample of the already present nodes of
fixed size $m$ is randomly chosen from the network which contains $t$ nodes. The
probability of chosing any node in the sample equals $m/t$. ii) Chosen nodes are
sorted by their degree in the decreasing order. For the nodes with
the same degree no additional rearangement is applied. iii) From such a sorted sample,
a new node is attached to the first $\omega$ nodes that have the
highest degree. %{\bf In the process of attachment of a new node it is possible If the number of nodes that have the same degree scheduled for
%picking $n(k)_{sched}$, is greater than the number of nodes that can still be chosen
%$\omega_{k}$, than the nodes are chosen with uniform equal probability
%$\omega_{k}/n(k)_{sched}$.} 
The third rule is a simple deterministic 
%\footnote{deterministic in a sense that no random selection is applied}
``winner takes it all" algorithm, which combined with the
first two rules produces very interesting macroscopic effects, as will be presented
in this paper. 

\begin{figure}
\bigskip
\includegraphics*[width=0.4\textwidth]{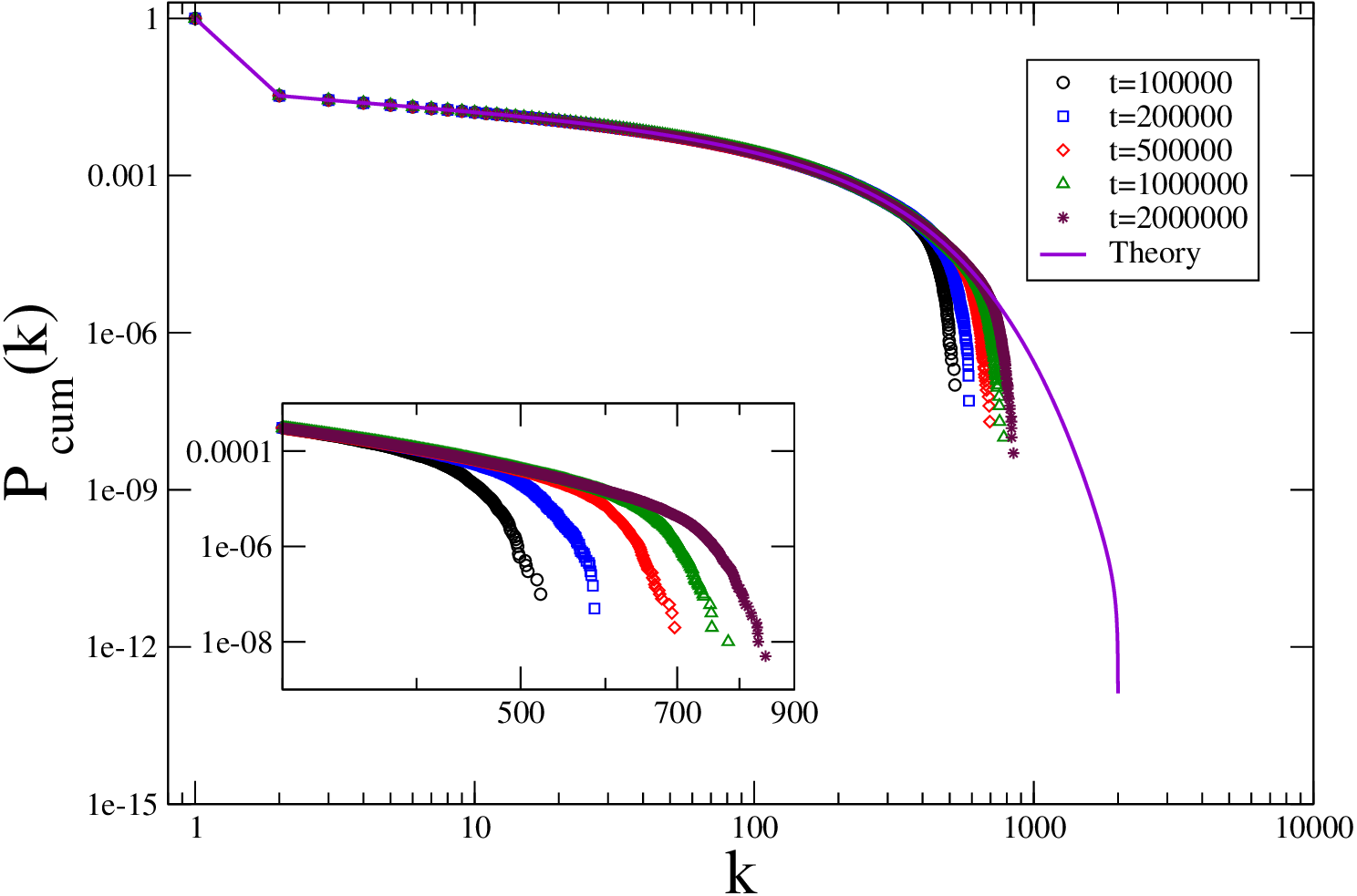}
\caption{\label{Fig: skaliranjezaclanak}Simulated cumulative probability functions
with $m=100$ and $\omega=1$ for different final network sizes $n_{max}$ are compared
to the theoreticaly obtained one. The figure clearly depicts the asymptotic approach of
simulation curves to the theoretical result. This implies that analytical results are
precise and that they sufficiently well describe the behavior of the
system when $n_{max} \rightarrow \infty$. The inlet gives the enlarged section with
the tails of the simulated cumulative probability distributions to better illustrate
the effects of the finite network size.}
\end{figure}

The nodes are numbered from 0, and the network is grown to the
size $n_{max}$. We averaged over 100 simulations for every investigated
$\omega$, $m$, and $n_{max}$ in order to get a statistically
relevant ensemble of network realizations. We also performed a
scaling investigation presented in Fig. \ref{Fig: skaliranjezaclanak} to see how
a simulated distribution behaves for different network sizes.

\section{Theory}

In the theoretical treatment of the node degree distribution we
decide to limit ourselves to the description of network with
$\omega = 1$. The reason for such an approach is a
cumbersome analytical study for the case of $\omega > 1$, which would include many
more summation terms that are analytically almost unentanglable. We
use the master equation approach of Dorogovtsev and Mendes
\cite{DMS00}. In this approach a new node enters the network at every moment $s$ and is
therefore denoted by $s$. It connects with one edge to the node with maximum degree
in the randomly selected sample of size $m$. Nodes in sample are selected from $t$
nodes that are already present in the network, so that every existing node has the
probability $\frac{m}{t}$ of entering the sample.

The probability that the node $s$ with degree $k$ will enter the sample of size $m$
at time $t$ and will be chosen for the attachment of the new node is  

\begin{eqnarray} \label{eq: v1}
v(k,m,t-1)&=&\sum_{l=0}^{m-1}
\left(\begin{array}{c}
\hat{B}(k,t-1) \\ l
\end{array}\right)
\nonumber \\
&\times&\frac{\left(\begin{array}{c}
N(k,t-1)-1 \\ m-l-1
\end{array}
\right)}{(m-l)\left(\begin{array}{c}
t \\ m
\end{array}\right)}.
\end{eqnarray}

Here the first binomial coefficient in the numerator represents number of possible
ways to chose $l$ nodes with degree smaller than $k$ into the sample, and

\beq\label{eq: B(k,t-1)}
\hat{B}(k,t-1)=\sum_{q=1}^{k-1}N(q,t-1),
\eeq

where $N(k,t)$ is the number of nodes with degree $k$ at time $t$.
The second binomial coefficient counts the number of possible ways to chose $m-1-l$
nodes with the same degree as node $s$ into the sample. 
This part of expresion (\ref{eq: v1}) accounts
for the possibility that in the selected sample exist other nodes with the same maximal
degree as $s$. Using the fact that $N(q,t)=P(q,t)\cdot
t$, together with approximation that for large $t$ one can approximate
$\left(\begin{array}{c} t \\ m \end{array}\right)$ with $t^m/m!$, we reduce the
expression (\ref{eq: v1}) to the following form:

\beq \label{eq: v2}
v(k,m,t-1)\simeq \frac{1}{t}\sum_{l=0}^{m-1}
\left(
\begin{array}{c}
m \\ l
\end{array}
\right)
\Pi(k,t-1)^{l}P(k,t-1)^{m-l-1},
\eeq

where

\beq \label{eq: Pi}
\Pi(k,t-1) \equiv \sum_{q=1}^{k-1} P(q,t-1).
\eeq

Using the well-established Dorogovtsev-Mendes master equation approach for
calculating the node degree distribution, for $k \geq 2$ we write

\begin{eqnarray}\label{eq: p(k,s,t)1}
p(k,s,t) &=& v(k-1,m,t-1)p(k-1,s,t-1) + \nonumber \\
&& \left(1-v(k,m,t-1)\right)p(k,s,t-1).
\end{eqnarray}

To calculate the probability distribution $P(k,t)$ that a randomly
chosen node has $k$ edges at time $t$, we average the probability
distribution of all nodes $s$, i.e. 

\beq \label{P(k,t)objasnjenje}
 P(k,t)=\frac{1}{t+1}\sum_{s=0}^t p(k,s,t).
\eeq

Thus we obtain

\begin{eqnarray}\label{eq: P(k,t)1}
P(k,t) &=& \frac{\zeta(k-1,t-1)}{t+1}P(k-1,t-1) + \nonumber \\
&& \left(\frac{t}{t+1}-\frac{\zeta(k,t-1)}{t+1}\right)P(k,t-1),
\end{eqnarray}

where

\beq \label{eq: zeta}
\zeta(k,t-1) \equiv \sum_{l=0}^{m-1}\left(
\begin{array}{c}
m \\
l \\
\end{array}
\right)
\Pi(k,t-1)^{l}P(k,t-1)^{m-l-1}.
\eeq

Assuming that Eq. (\ref{eq: P(k,t)1}) has a stable asymptotic solution for $t\gg 1$
thus changing the time-dependent probability distribution into time independent
$P(k,t)=P(k)$, we obtain the following closed form:

\beq \label{eq: P(k)1}
 P(k) = \zeta(k-1)P(k-1)-\zeta(k)P(k).
 \eeq

Equations (\ref{eq: P(k)1}) are a polynomials of order $m$ and hold for all $k \geq
2$. Written as polynomials, they adopt the following form:

\begin{eqnarray} \label{eq: P(k)2}
a(0)P(k)^m + a(1)P(k)^{m-1} + ... + a(l)P(k)^{m-l} &+&...\nonumber\\
+ (1+a(m-1))P(k)&-&  \nonumber\\
\sum_{l'=0}^{m-1}\left(
\begin{array}{c}
m \\
l' \\
\end{array}
\right)\left(\sum_{q=1}^{k-2}P(q)\right)^{l'}P(k-1)^{m-l'} &=& 0, \nonumber\\
\end{eqnarray}

where the coefficients $a(l)$ are

\beq\label{eq: a(l)}
a(l)=\left(\begin{array}{c}
m \\
l \\
\end{array}\right)
\left(\sum_{q=1}^{k-1}P(q)\right)^{l}.
\eeq

For theoretical treatment of $P(1)$ as our boundary condition the following equation
holds:

\beq \label{eq: p(1,s,t)1}
p(1,s,t) = \delta_{s,t}+(1-\delta_{s,t})\left(1-v(1,m,t-1)\right)p(1,s,t-1),
\eeq

with an obvious relation for probability that a node with one edge at time $t-1$
will adopt a new edge at time $t$:

\beq \label{eq: v(1,s,t-1)}
v(1,s,t-1)=\frac{P(1,t-1)^{m-1}}{t}.
\eeq

Using a procedure similar to that already mentioned above, we
obtain the asymptotic value for $P(1)$:

\beq \label{eq: P(1)}
P(1) = 1-P(1)^m.
\eeq

\begin{figure}
\bigskip
\includegraphics*[width=0.4\textwidth]{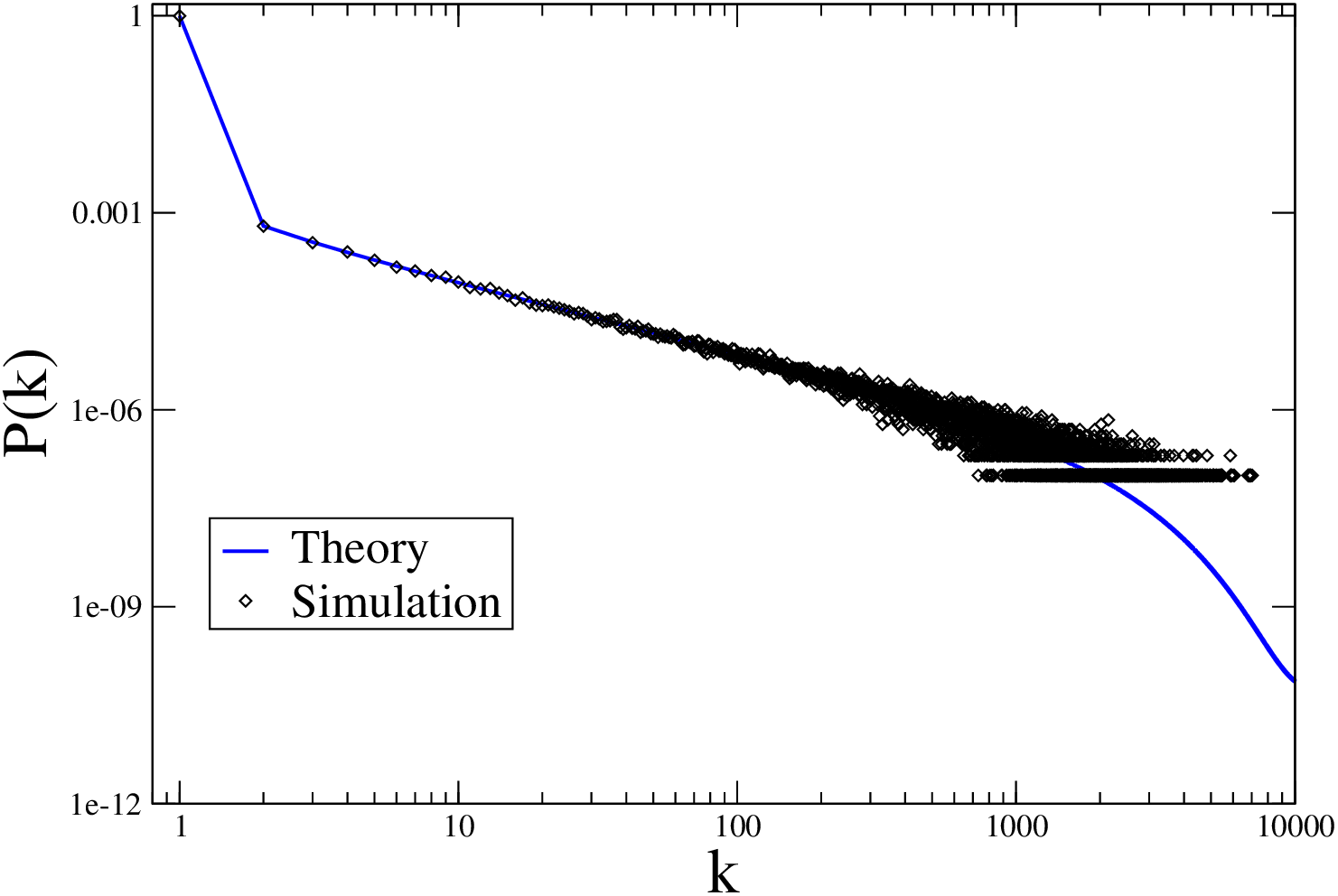}
\caption{\label{Fig: scattering}Theoretical probability distribution (solid line)
nicelly follows simulation data (black diamonds) for $m=1000$.
Scattering in the tail is a consequence of low probability
fluctuations induced by finite size effects. The reader should also
note a big jump of probability for $P(k=1)$.} 
\end{figure}

Unfortunately, the set of Eqs. (\ref{eq: P(k)2}) and (\ref{eq: P(1)}) is
analytically unsolvable and is therefore solved numerically.
The solutions of these polynomial equations show excellent agreement with
numerical simulations as can be seen in Figs. \ref{Fig: scattering}, \ref{Fig:
SimTheoCM10}, \ref{Fig: SimTheoCM100}, and \ref{Fig: SimTheoCM1000}. These findings further
vindicate the master equation approach followed in this paper.

%Using the same technique it is possible to find equations for
%$\omega \geq 2$. Obtained equations are cumbersome for calculation
%for large values of $m$ and $\omega$ and are therefore manageable
%only for small values. The equation for $k=\omega$ is:

%\beq \label{eq: P(omega,k=omega)}
%P(k=\omega)=1-\sum_{z=0}^{\omega - 1}\left(
%A(\omega)^z P(\omega)^{m-z}(\omega-z)\left(m-z,z\right)\right).
%\eeq

%Here $\left(m-z,z\right)$ represents the multinomial coefficient,and we have made the
%substitution:

%\beq\label{eq: A(k,x)} 
%A(k)\equiv
%\sum_{q=k+1}^{\infty}P(q). 
%\eeq

%The equation for general $k > \omega$ is

%\beq\label{eq: P(omega,k)}
%P(k) = \frac{\Pi(k-1)P(k-1)}{1+\Pi(k)}.
%\eeq

%Here $\Pi(k)$ stands for

%\begin{eqnarray}\label{eq: Pi(k)}
%\Pi(k)&=& \sum_{z=0}^{\omega-1}A(k)^z\sum_{y=0}^{m-1-z}P(k)^y
%B(k)^{m-1-y-z}\nonumber \\
%&\times&\left(z,y+1,m-z-y-1\right)G(y,z).
%\end{eqnarray}

%The term $\left(z,y+1,m-z-y-1\right)$ again represents the multinomial coefficient
%and $B(k)$ and $G(y,z)$ are

%\beq\label{eq: B(k)}
%B(k)\equiv \sum_{q=\omega}^{k-1}P(q),
%\eeq

%\beq\label{eq: G(y,z)}
%G(y,z)=(\omega-z)\Theta(z-\omega+y+1)+\Theta(\omega-z-y-1).
%\eeq

%Obviously, $\Theta(q)$ represents a Heaviside step function.

\section{Discussion}

As we have mentioned in the preceding section, a master equation
approach yields a chain of the polynomial equations (\ref{eq: P(k)2}) and
(\ref{eq: P(1)}). Note the fact that $P(k^{*})$ representing the
probability that a randomly chosen node will have a degree $k^{*}$
depends only on degree probabilities that are equal or less than
$k^{*}$ \ref{eq: P(k)2}. We have calculated the roots of the system to get a degree
probability distribution. 

\begin{figure}
\bigskip
\includegraphics*[width=0.4\textwidth]{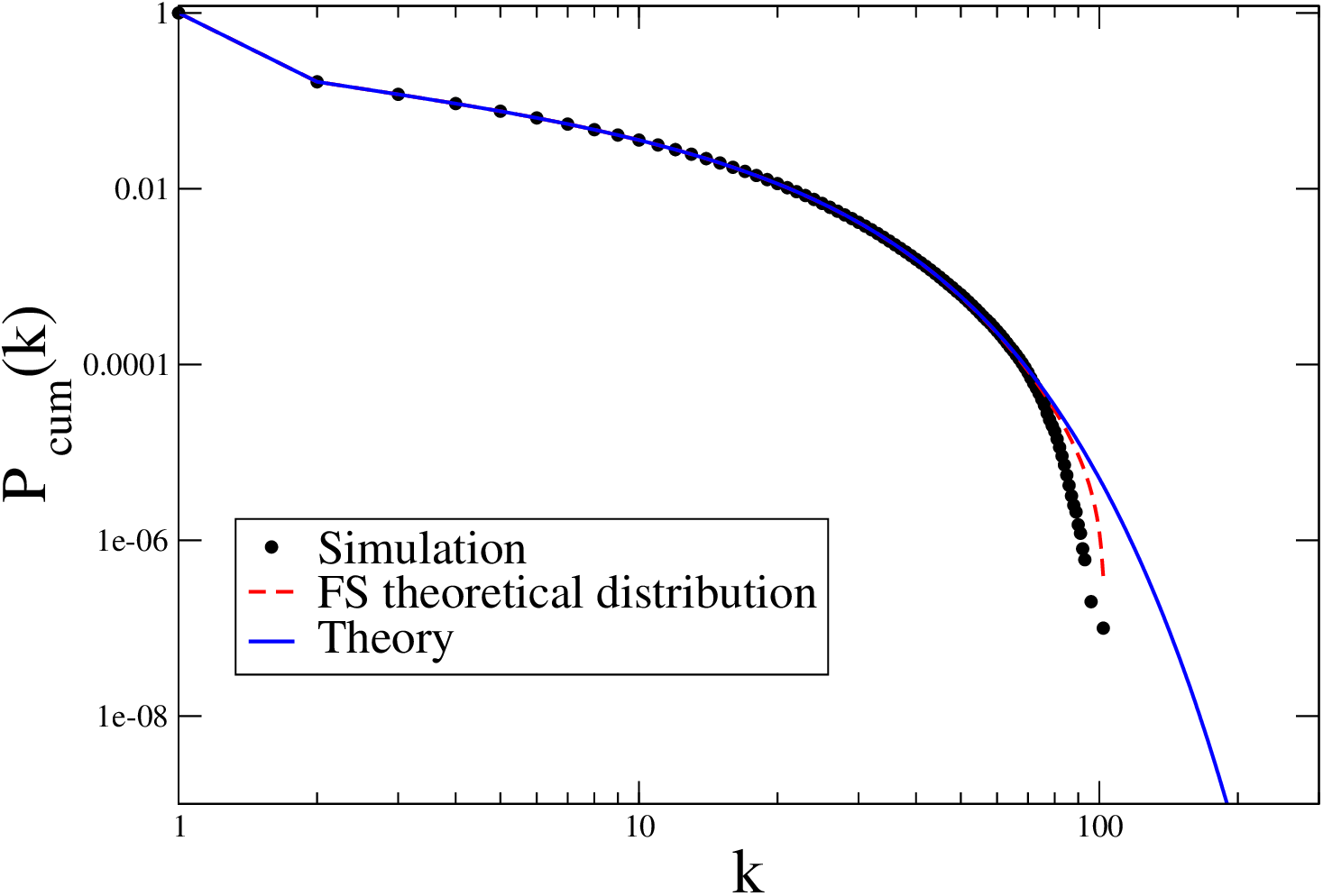}
\caption{\label{Fig: SimTheoCM10}For $m=10$ the theoretical distribution (solid line)
nicelly follows simulation data (black dots). The disagreement in the very tail is
explained by finite size effects of simulated data. However, a FS
theoretical distribution obtained by transformation (\ref{eq: renormala}) shows even
better agreement with simulation} 
\end{figure}

All simulated data and analytical roots of polynomial equations
exhibit a big jump from $P(k=1)$ to $P(k=2)$ of order of a magnitude
or more. The difference $P(k=1)-P(k=2)$ depends strongly on the
size of a chosen sample $m$. If the size of the sample is larger,
then there is higher probability that a node of degree larger
than $1$ will enter the sample, and collect the new link. The
smaller the sample the greater the probability that only
nodes of degree one will be chosen in the sample, thus lowering the
overall amount of nodes of degree one. The obtained analytical
solutions from Eq. (\ref{eq: P(1)}) are in excellent agreement with
simulational results regarding to this jump. The average relative error for
$m\in\{10, 100, 1000\}$ simulation and theory is $4.3\cdot 10^{-5}$, and gets smaller
as the sample size $m$ grows larger for $n_{max}=10^{6}$. 

\begin{figure}
\bigskip
\includegraphics*[width=0.4\textwidth]{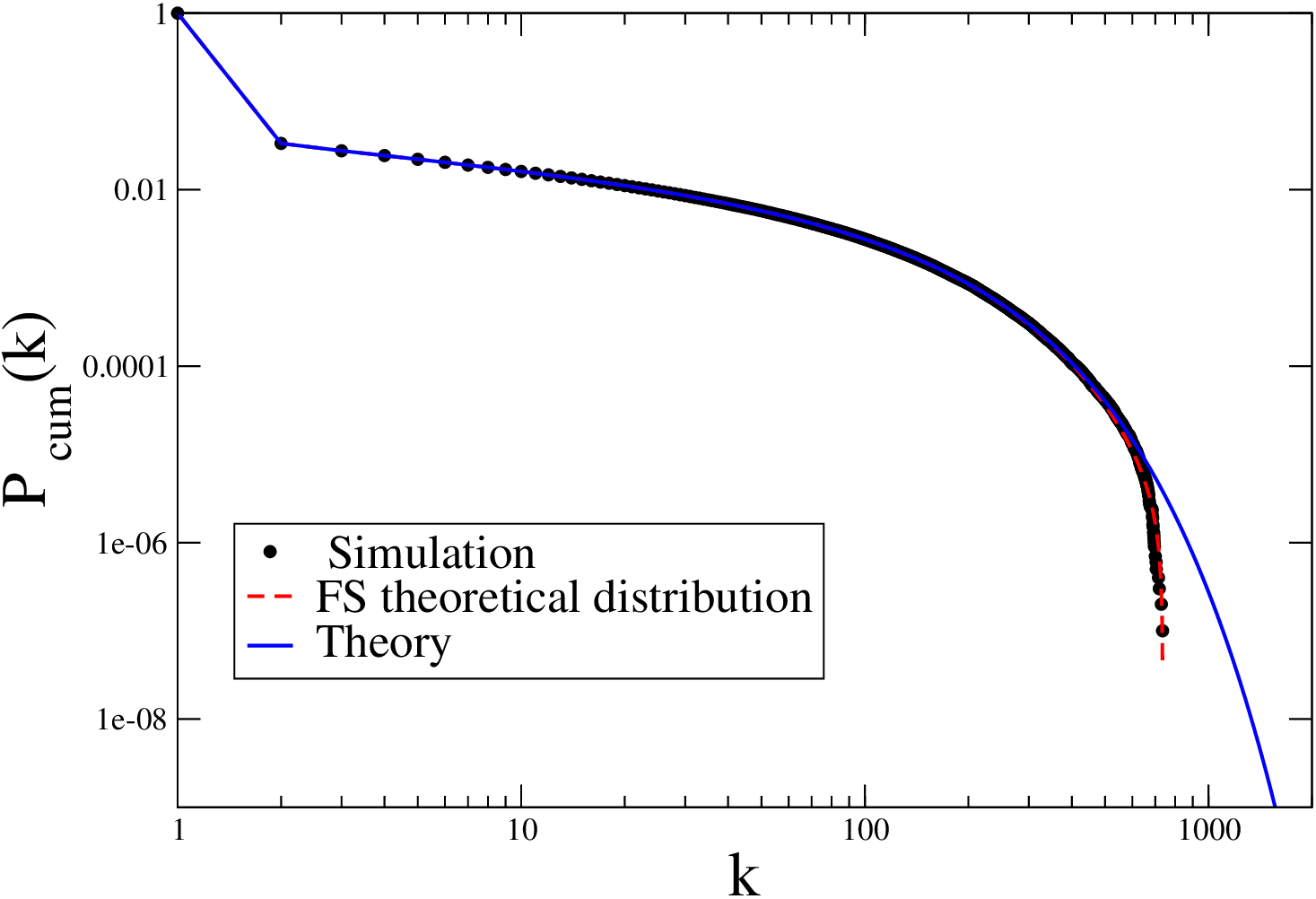}
\caption{\label{Fig: SimTheoCM100}For $m=100$ it is easy to see that the theoretical
distribution follows simulational data very well, and FS theoretical distribution
even better.}  
\end{figure}

All simulated data exhibit a strong scattering in the tail. The
scattering is a consequence of low probability fluctuations and makes the comparison
between theory and simulation more dificult, Fig.\ref{Fig: scattering}. In order to
straighten up the data and compare theory and simulation, it is possible to use
exponential binning or to transform probability distribution into the cumulative
probability distribution. We implemented the second approach and produced a cumulative
degree probability distribution $P_{cum}$. 

\beq\label{eq: P_cum} 
P_{cum}(k)=\sum_{q=k}^{\infty}P(q).
\eeq

\begin{figure}
\bigskip
\includegraphics*[width=0.4\textwidth]{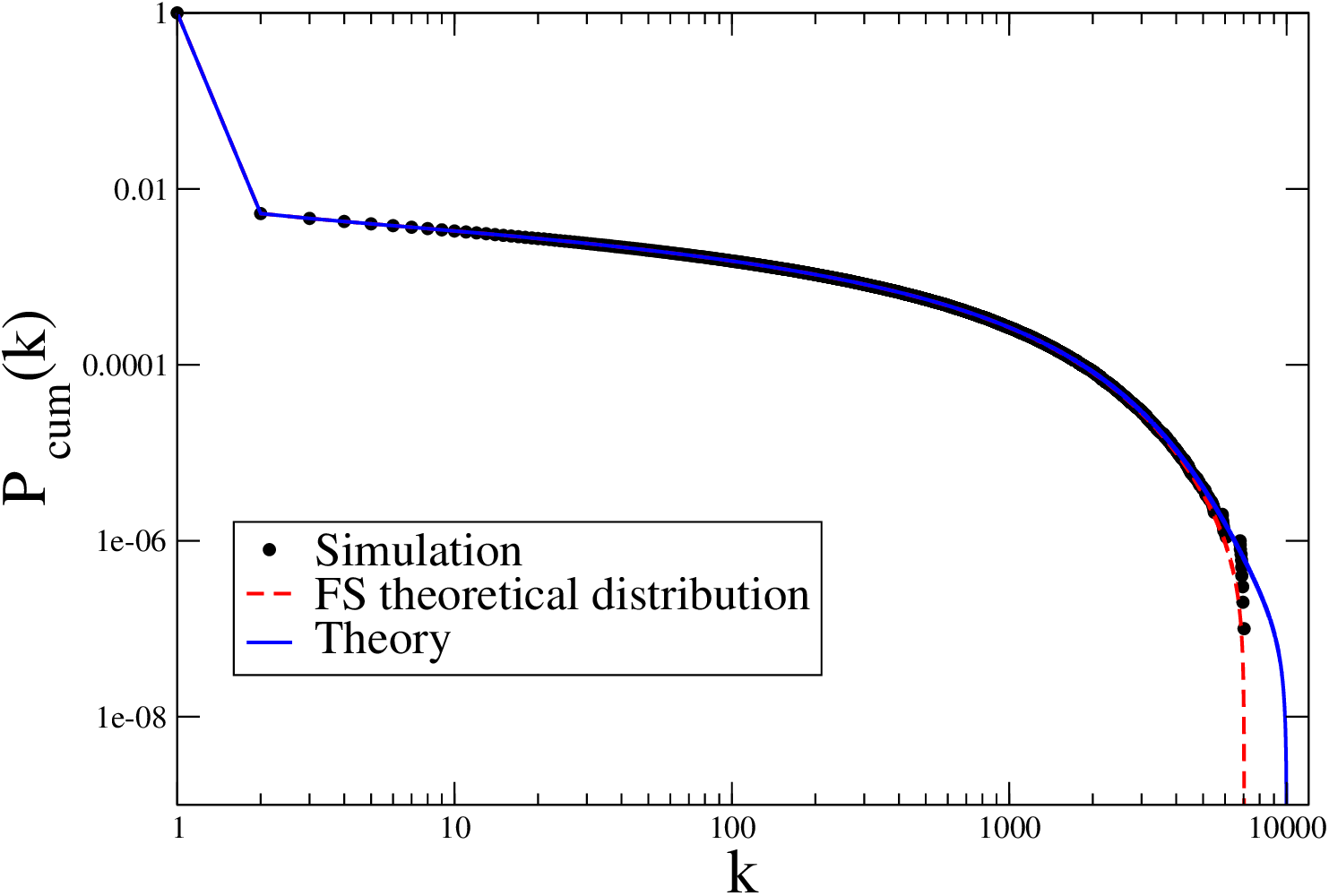}
\caption{\label{Fig: SimTheoCM1000}$m=1000$ is the largest monitored sample size but
is still small enough compared to simulational number of nodes
$n_{max}=10^6$. Theory is in excellent agreement with simulation. } 
\end{figure}

This distribution contains the same system information as the degree
distribution, but is much smoother in the tail. We compared our
theoretical curve with the simulated one and found an excellent match between
theory and simulations. The results of the comparison between
simulation and theory are presented in Figs.
\ref{Fig: SimTheoCM10}, \ref{Fig: SimTheoCM100}, and \ref{Fig: SimTheoCM1000}. The
relative disagreement observed in the tails is a consequence of finite size
effects, Fig.\ref{Fig: skaliranjezaclanak}. 
%Every simulated distribution has the
%largest observed degree in the network, $k_{max}\leq n_{max}$. 
Since our theoretical curve  falls down relatively slowly, as can be seen in table
\ref{Tab: tabela}, the summation of probabilities for $k>k_{max}$ in Eq. (\ref{eq:
P_cum}) contributes strongly to the cumulative degree probability in
the tail. To get an even better match, we calculated ``renormalized"
cumulative probability distribution 

\beq\label{eq: renormala}
\tilde{P}_{cum}(k)=\frac{\sum_{q=k}^{k_{max}}P(q)}{1-\sum_{q=k_{max}+1}^{\infty}P(q)}.
\eeq

\begin{figure}
\bigskip
\includegraphics*[width=0.4\textwidth]{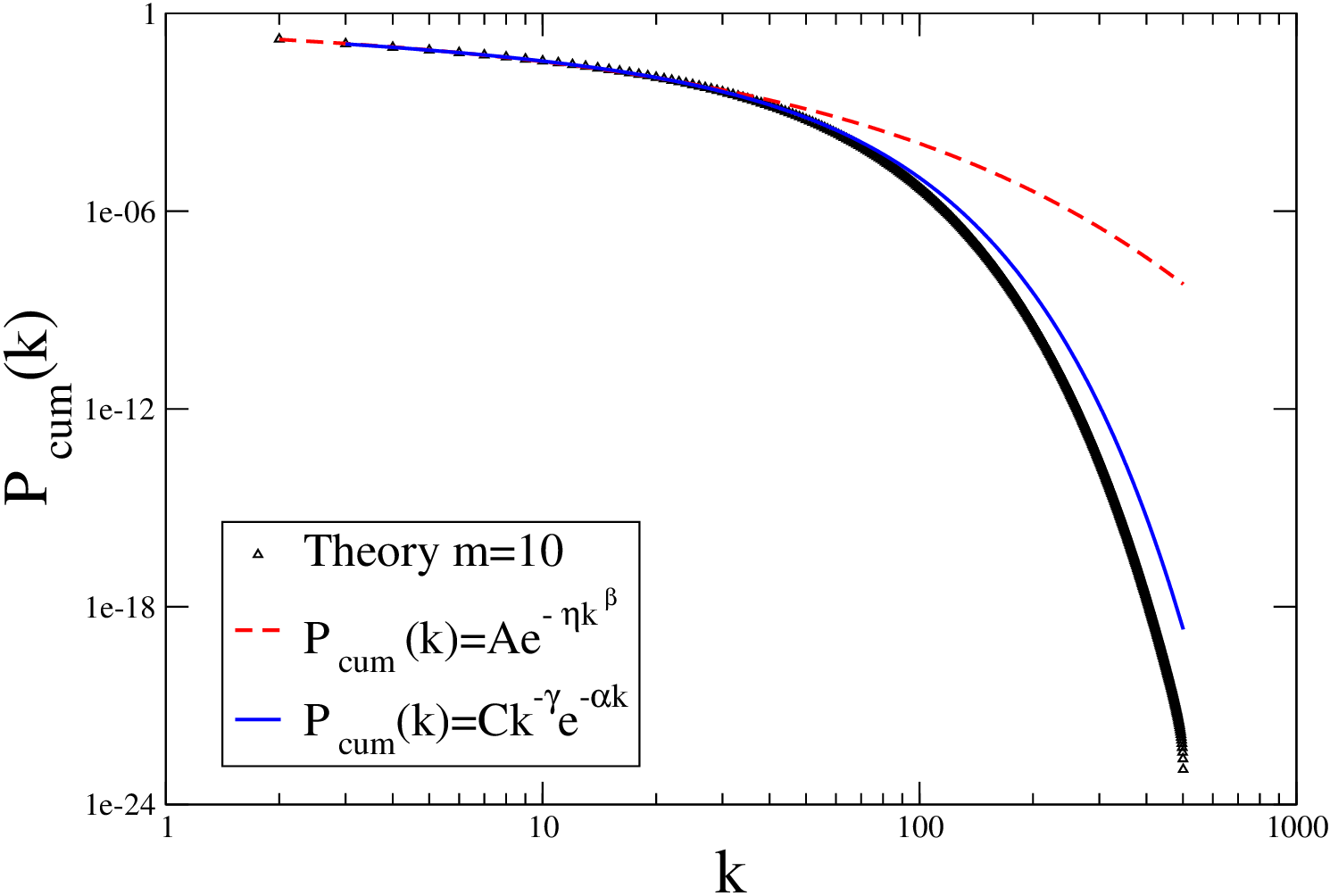}
\caption{\label{Fig: FitCM10ZaClanak}Two different functions: i) stretched
exponential and ii) power law with the cut off are fitted on theoretical data for
sample size $m=10$. This figure clearly shows that the power law with
the exponential cut off better describes the tail of the
theoretical distribution. } 
\end{figure}

This Finite Size cumulative probability distribution (hereafter denoted as FS
theoretical distribution) is even better in describing
finite size effects, as shown in the Figs.\ref{Fig: SimTheoCM10}, \ref{Fig:
SimTheoCM100}, and \ref{Fig: SimTheoCM1000}. 

To obtain a description of the degree distribution in the thermodynamical limit, we
fitted theoretical cumulative degree distribution (theoretical and not
simulational distribution was also used since it does not suffer from finite size
effects) with the stretched exponential
(\ref{eq: Stretched}) and power-law distribution with the exponential cut off
(\ref{eq: nas}), \cite{MI04a}: 

\beq\label{eq: Stretched}
P_{cum}\sim e^{-\eta k^{\beta}},
\eeq

\beq\label{eq: nas}
P_{cum}\sim k^{-\gamma}e^{-\alpha k}.
\eeq

\begin{figure}
\bigskip
\includegraphics*[width=0.4\textwidth]{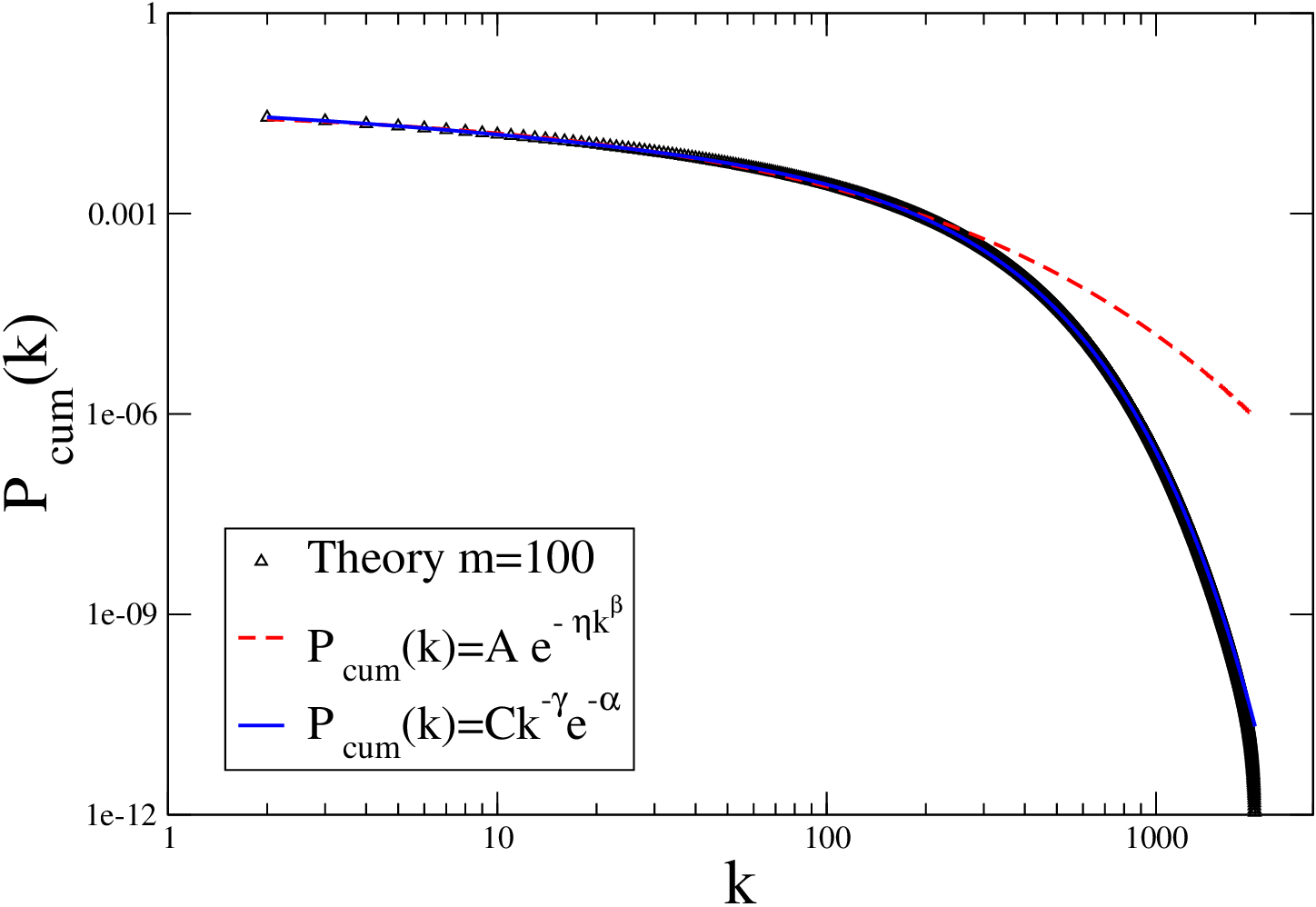}
\caption{\label{Fig: FitCM100ZaClanak}Fitted curves for $m=100$. Power law with the
exponential cut off represents the theoretical distribution very well.}

\end{figure}

For fitting purposes we used all theoretical $P_{cum}(k)$ values, except
$P_{cum}(1)$, because its value is clearly not determined by the scale-free-like
behavior as opposed to all other $k$ values. 
Both distributions fit our overall results very well, as
presented in Table \ref{Tab: tabela}, and Figs.
\ref{Fig: FitCM10ZaClanak}, \ref{Fig: FitCM100ZaClanak}, and \ref{Fig:
FitCM1000ZaClanak}. The correlation coefficients of the
fitted distributions are all above 0.99 margin, proving that both
fitting models are capable of describing theoretically obtained curves very well.
The power law with the exponential cut off always has just a slightly higher
correlation coefficients than stretched exponential for the same sample size
$m$. Figures clearly show that the reason for this behavior is
much better description of the tail, which power law with
exponential cut off exhibits. Stretched exponential is clearly not suitable for the
description of the tail properties.

It is worthy to mention that the power-law distribution with
the exponential cut off has already been obtained in a similar model
\cite{MI04a}, which has shown that exponential parameter $\alpha$ is trivially
connected with the sample size $m$ by the relation
$\alpha=\frac{1}{m}$. Although one cannot expect this relation to be valid for this
model also, the parameter $\alpha$ is very close to $1/m$, and this coincidence
is better for larger $m$, as can be seen in table \ref{Tab: tabela}. In our opinion,
it would be interesting to measure $\alpha$ in some 
observed network distributions of a similar shape and compare it
with the expected sizes of samples on which the new node has the possibility
of creating a link.

\begin{figure}
\bigskip
\includegraphics*[width=0.4\textwidth]{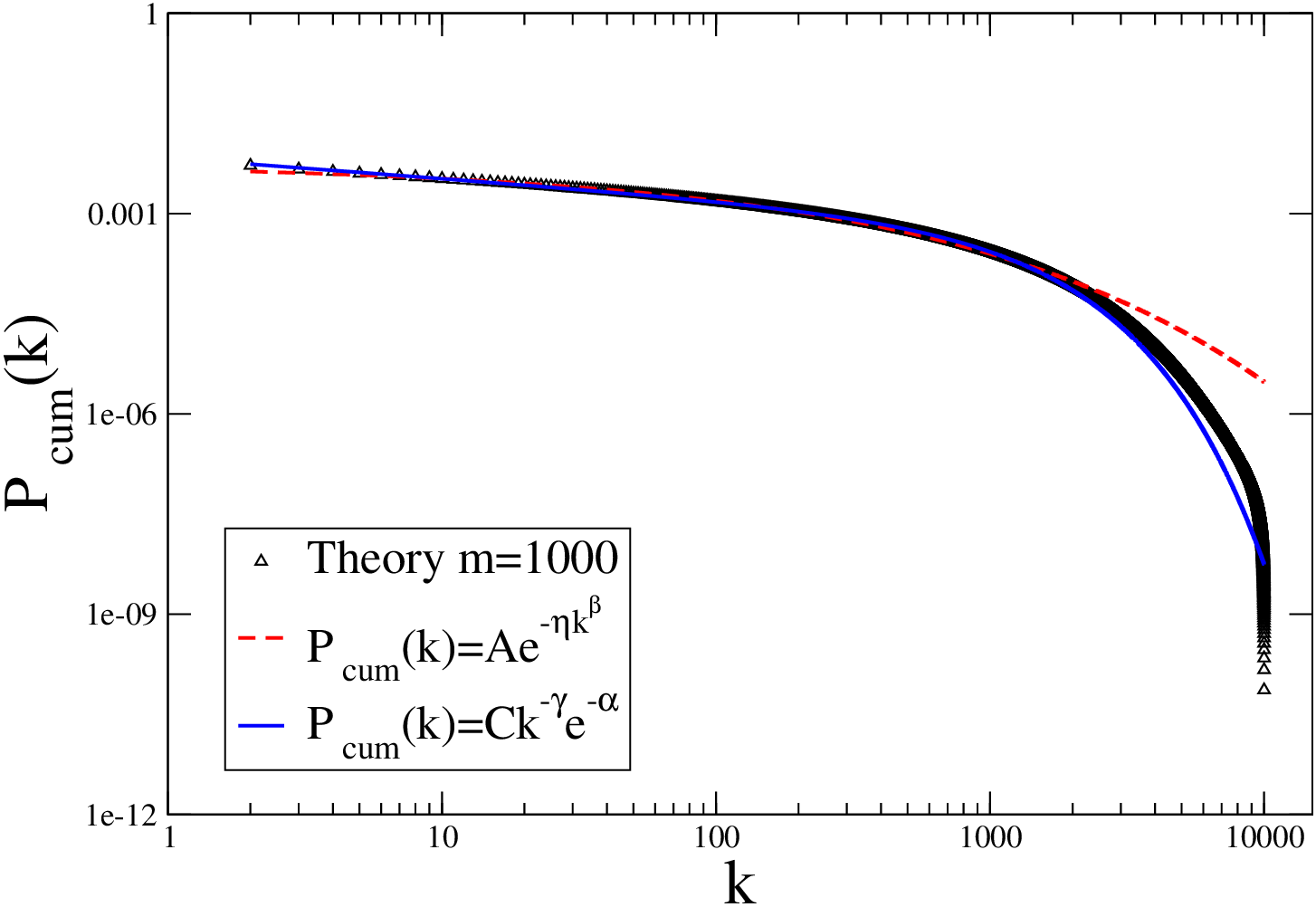}
\caption{\label{Fig: FitCM1000ZaClanak}Excellent agreement of fitted and theoretical
distributions for $m=1000$.}
\end{figure}

\small
\begin{table}[h]
\begin{center}
\begin{tabular}{|l||c|c|c||c|c|c|}
\hline
&\multicolumn{3}{c|}{}&\multicolumn{3}{c|}{}\\
&\multicolumn{3}{c|}{$P_{cum}\sim k^{-\gamma}e^{-\alpha k}$}&\multicolumn{3}{c|}{$
P_{cum}\sim e^{-\eta k^{\beta}}$}\\
\cline{2-7}
&$\gamma$&$\alpha$&{\bf corr}&$\eta$&$\beta$&{\bf corr}\\
\hline\hline
m=10&0.5501&0.0765&0.9980&0.9829&0.4718&0.9976\\
m=100&0.4026&0.0092&0.9995&0.3894&0.4385&0.9987\\
m=1000&0.3067&0.0011&0.9994&0.2230&0.3823&0.9978\\
\hline
\end{tabular}
 \caption{\label{Tab: tabela} Fitted distribution parameters
for different sample sizes. Correlation coefficients show
excellent agreement between the theoretical distribution data and the presented fits.}
\end{center}
\end{table}
\normalsize

\begin{figure}
\bigskip
\includegraphics*[width=0.4\textwidth]{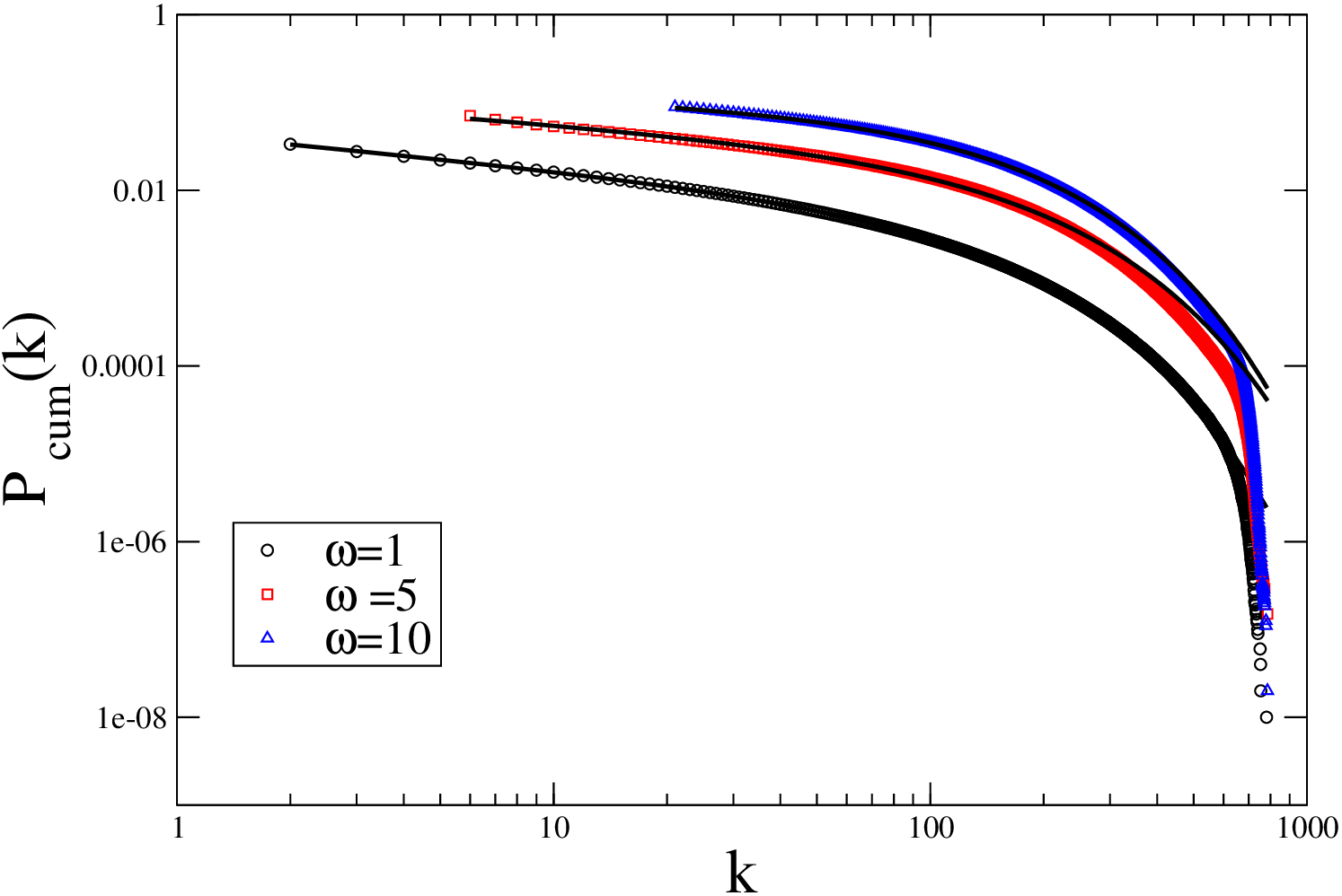}
\caption{\label{Fig: manylinx}Evidence that the distributions for $\omega
> 1$ fall in the same class as the distributions studied
analytically. The situation with $m=100$ is presented.} 
\end{figure}

Finally, let us briefly discuss simulational results for $\omega > 1$. Simulational
results for the cumulative probabability distribution (without $P(\omega)$) are displayed 
in Fig. \ref{Fig: manylinx}. The typical characteristics of the distribution are
equivalent to the $\omega = 1$ case. The degree $k=\omega$ has substantially larger
probability compared to all other degrees. The cumulative probability distribution
for $k \geq 2$ obtained in the simulations displays the scale-free-like properties. 
These simulational distributions can be well fitted with the power
law with the exponential cut off (\ref{eq: nas}) as shown in Fig. \ref{Fig:
manylinx}.  

\section{Conclusion}

We have shown that using a simple ``winner takes it all" algorithm,
together with the fact that nodes do not possesses complete information on
network structure, a macroscopic node-degree power law is created.
Evolution of real networks is still an open question and we have
shown that realistic imperfect knowledge can have a substantial
effect on network growth. Although the field of complex networks
has made great progress during the last few years, there is still much open
space for research of microscopic models that describe the
formation of complex networks with certain expected features. Our
results clearly show that stohastic-deterministic processes even as simple
as that described in this paper can be used to reproduce some
macroscopic effects of complex networks. Moreover, in this paper
as well as in \cite{MI04a}, we have demonstrated that the power law
with the exponential cut off can be a significant distribution for
types of networks in which information filtering is performed. New
findings in social contact networks \cite{Batorski04} lead us to
believe that the power law with the exponential cut off and stretched
exponentials should be studied more intensively in the future.

{\bf Acknowledgment.} This work was supported by the Ministry of Science and
Technology of the Republic of Croatia under the contract numbers 0098002 and 0098004.

 %------------------staro---------------------------------------------

%\beq P(k) = \frac{m\Pi(k-2)^{m-1}P(k-1)}{1+m\Pi(k-1)^{m-1}} ; \eeq

%\begin{eqnarray}
 %P(k) & = & m\left(\Pi(k)-P(k)-P(k-1)\right)^{m-1}P(k-1) - \nonumber \\
 %&& m\left(\Pi(k)-P(k)\right)^{m-1}P(k) ;
 %\end{eqnarray}

 %\beq \left(\Pi(k)-P(k)\right)^{m-1}\simeq
 %\Pi(k)^{m-1}-(m-1)\Pi(k)^{m-2}P(k); \eeq

 %\beq \left(\Pi(k)-P(k)-P(k-1)\right)^{m-1} \simeq
 %\Pi(k)^{m-1}-(m-1)\Pi(k)^{m-2}(P(k)+P(k-1) \eeq

 %\begin{eqnarray}
  %P(k) &=& m\left(\Pi(k)^{m-1}-(m-1)\Pi(k)^{m-2}(P(k)+P(k-1))\right)P(k-1) -\nonumber \\
    %&& m\left(\Pi(k)^{m-1}-(m-1)\Pi(k)^{m-2}P(k)\right)P(k);
 %\end{eqnarray}

%\beq P(k) = m\Pi(k)^{m-1}P(k-1)-m\Pi(k)^{m-1}P(k); \eeq

%\beq P(k)= - m\Pi(k)^{m-1}\frac{dP}{dk}; \eeq

%\beq \Pi(k)= \left(-\frac{P}{mP'}\right)^{\frac{1}{m-1}}; \eeq

%\beq (m-1)\Pi(k)^{m-2}P(k)=
%-\frac{1}{m}-\frac{P}{m}\left(\frac{P''}{(P')^2}\right); \eeq

%\beq
%(m-1)\left(-\frac{P}{mP'}\right)^{\frac{m-2}{m-1}}=-\frac{1}{m}+\frac{1}{m}\frac{PP''}{(P')^2};
%\eeq

%\beq PP''-(P')^2+(m-1)PP'=0 \eeq

%\beq P(k)=Ce^{\frac{\alpha}{m-1}e^{(-(m-1)k)}}; \eeq

%--------------------------------------------------------------

\end{document}